\documentclass[a4paper]{article}

\usepackage{indentfirst}
\usepackage{epsfig}

\begin{document}

\title{Commissioning of the Silicon Strip Detector (SSD) of ALICE}


\author{Panos Christakoglou\footnote{The author acknowledges the valuable contributions of Marek Chojnacki, Andrea Dainese, Marco van Leeuwen, Francesco Prino and Andrea Rossi. Special thanks go to the ALICE SSD Collaboration.} for the ALICE Collaboration\\\\
        NIKHEF - Utrecht University\\
        E-mail: Panos.Christakoglou@cern.ch}
\maketitle


\begin{abstract}
The latest results from the commissioning of the SSD with cosmics 
are presented in this paper. The hardware status of the detector, the front-end 
electronics, cooling, data acquisition and issues related to the on-line 
monitoring are shown. In addition, the procedures implemented and followed to 
address the alignment with the rest of the ITS sub-detectors along with both 
on-line and off-line calibration strategies are described. Finally, results 
from simulations as well as from the reconstruction of cosmic data 
demonstrating the performance of the detector are presented, proving that the 
SSD is ready for the forthcoming proton-proton data taking.
\end{abstract}



\section{Introduction}

The Inner Tracking System (ITS) of the ALICE experiment \cite{Ref:Alice}, 
consists of six cylindrical layers of silicon detectors, the Silicon Pixel 
Detectors (SPD), the Silicon Drift Detectors (SDD) and the Silicon Strip 
Detectors (SSD). The outer layers are made of double sided Silicon Strip 
Detectors mounted on carbon-fiber support structures \cite{Ref:Alice}. The 
SSD is crucial for the connection of tracks from the main tracking device 
of ALICE, the Time Projection Chamber (TPC), to the ITS and also provides 
dE/dx information to assist particle identification for low-momentum particles
\cite{Ref:AlicePPR}. The detector consists of 1698 modules each one having 
768 P- and 768 N-side strips, resulting in total to more than 2.6 million 
channels. The SSD has been actively participating in all the commissioning 
and run activities as well as in all the data taking periods of the ALICE 
experiment.

\section{Commissioning results}
The installation of the entire ITS in its final position took place in June 
2007. During the first commissioning phase (July - October 2007), all the 
connections were checked in detail. In December 2007 and in February/March 
2008, the SSD participated in the first and second cosmics run respectively, 
during which partial cooling was available. As a consequence only a small 
fraction of the detector was included in the data acquisition system. The 
installation of the services and the upgrade of the cooling plant took place 
in May 2008, allowing us to include all the SSD modules in the third cosmics 
run that started in June 2008. The scope of this run was to collect a suitable 
data sample to perform the first part of the detector's alignment and charge 
calibration. The following paragraphs will summarize the results obtained 
during the summer of 2008.

\subsection{Detector operation}

\begin{figure} 
\includegraphics[width=7cm,height=6cm]{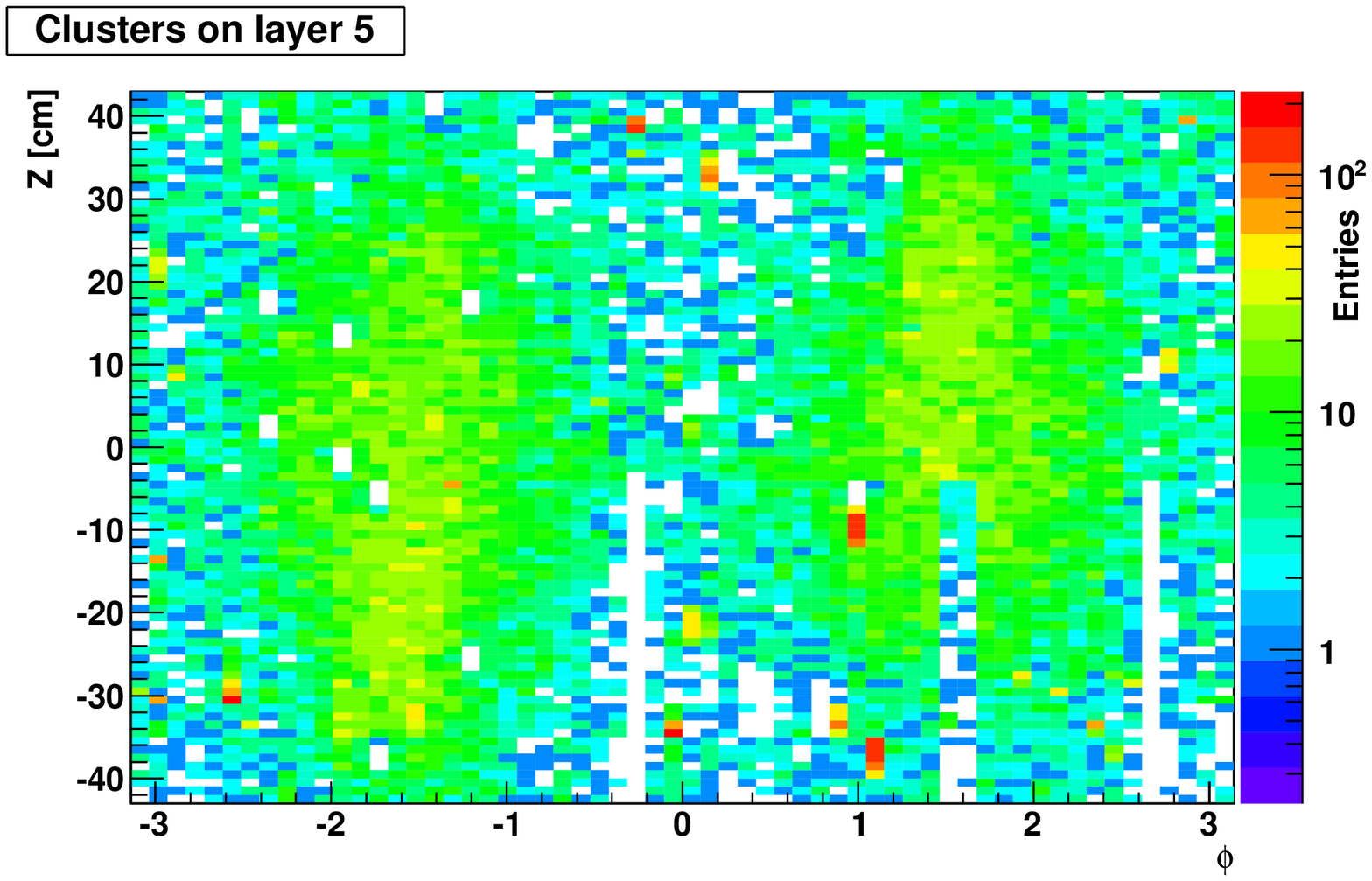} 
\includegraphics[width=7cm,height=6cm]{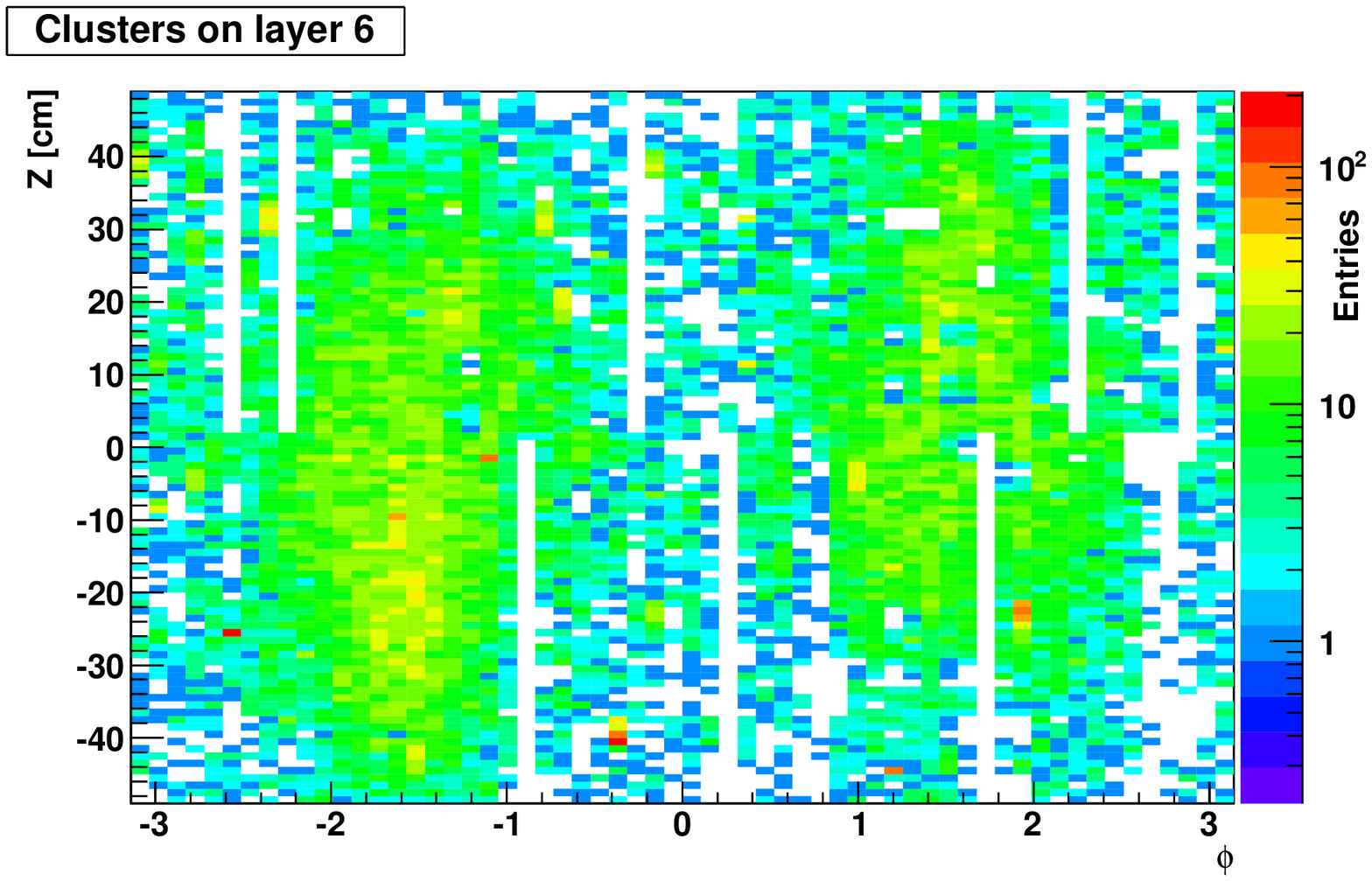} 
\caption{The distribution of reconstructed clusters from the cosmic data 
taking for layer 5 (left) and layer 6 (right) as a function of the 
azimuthal angle $\phi$ and the global z coordinate (along the beam axis).} 
\label{fig:cosmics} 
\end{figure}

During the cosmics run 1477 out of 1698 SSD modules took data in summer 2008. 
The fraction of bad strips was $\approx1.5\%$.  Most of the modules not 
included in the data taking were drawing unexpected high bias current and 
were switched off as a precaution, pending further investigation, although 
their performance was still good. The resulting signal over noise ratio was 
better than $S/N > 40$. Figure \ref{fig:cosmics} shows the distribution in z 
(along the beam axis) and $\phi$ (azimuthal angle) of the reconstructed 
clusters for both SSD layers. The SPD FastOR \cite{Ref:Alice} was used as a 
trigger detector, the rate of which reached the value of $\approx 0.18$ Hz.

\subsection{Alignment}

\begin{figure} 
\includegraphics[width=4.8cm,height=6cm]{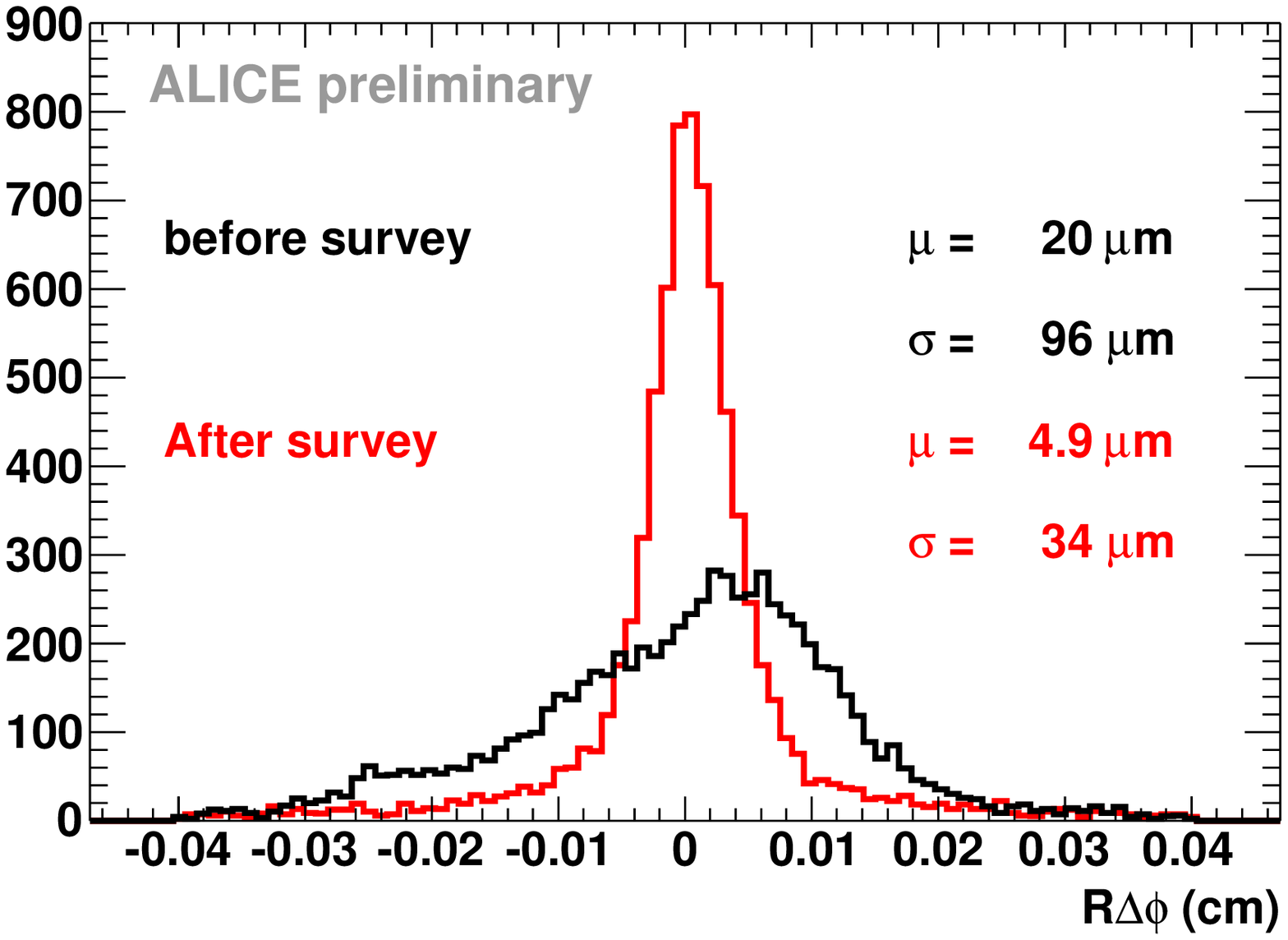} 
\includegraphics[width=9.6cm,height=6cm]{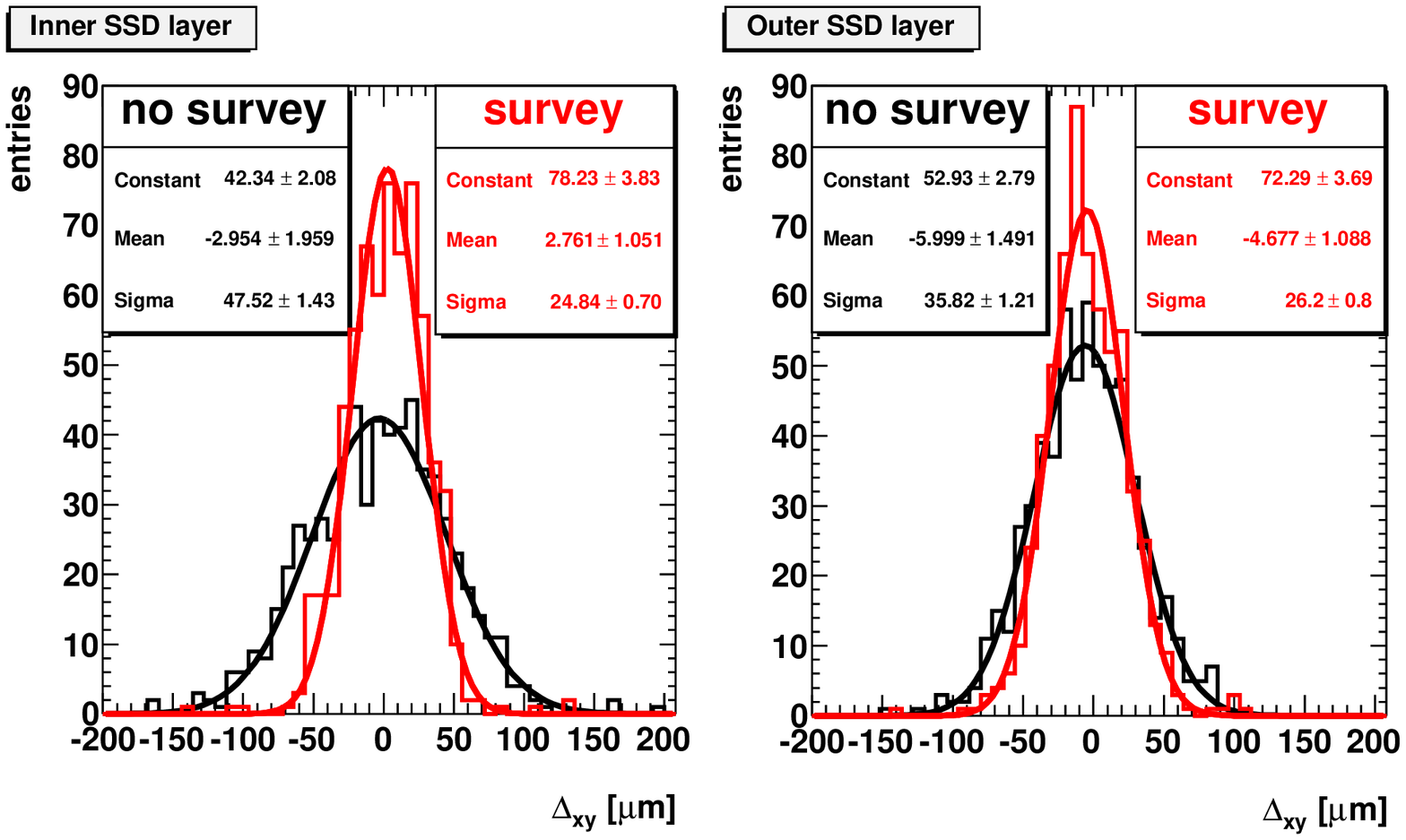}
\caption{Track to point residuals with and without the SSD survey information 
(left plot). Track to extra cluster $\Delta xy$ for the two SSD layers (middle 
plot: layer 5, right plot: layer 6) before and after the usage of the SSD 
survey data.} 
\label{fig:alignment} 
\end{figure}

The study of the displacements and deformations of the SSD modules enhances 
the knowledge of the realistic detector geometry and thus contributes to the 
tracking performance. This is performed using cosmic data as well as pp 
events when the latter will be available. The starting point of the alignment 
procedure is the optical measurement (survey) performed during the 
construction. Then different tools are used in order to extract the relevant 
information. This allows us to align sensitive elements mounted on common 
mechanical supports (in the SSD case these are the 72 ladders), whereas 
with higher statistics we will be able to move to the level of a single module.
Figure \ref{fig:alignment}-left shows the distributions of the distances in 
$r-\phi$ between the track fitted on the outer SSD layer and the points 
measured in the inner SSD layer. The inclusion of the survey data reduces 
significantly not only the mean of the distributions (from $20 \mu m$ to 
$4.9 \mu m$) but also its spread (from $96 \mu m$ to $34 \mu m$). In addition, 
the middle and right plots of fig. \ref{fig:alignment} show the track to point 
distance for the SSD extra clusters in the $r\phi$ plane without and with the 
inclusion of the survey data for the inner and outer SSD layers respectively. 
The extra clusters are created when particles cross a region with an acceptance 
overlap between two adjacent modules. In both cases, the inclusion of the SSD 
survey data improves the extracted resolution by a factor of $\approx 1.9$ and 
$\approx 1.3$ for the two layers \cite{Ref:AlignmentNote}.

\subsection{Charge calibration}

The gain calibration of the SSD has two components: relative calibration of 
the P and N sides and overall calibration of ADC values to energy loss. The 
charge matching is a strong point of double sided silicon sensors and helps 
to remove fake clusters. Both calibrations relied on cosmics. Already in the 
laboratory the calibration constants were determined, using cosmics on a spare 
ladder using a data-acquisition system setup which matches the one used in 
the experiment. These constants were refined during the cosmics runs with the 
pixel trigger. 

\begin{figure} 
\includegraphics[width=3.9cm,height=6cm]{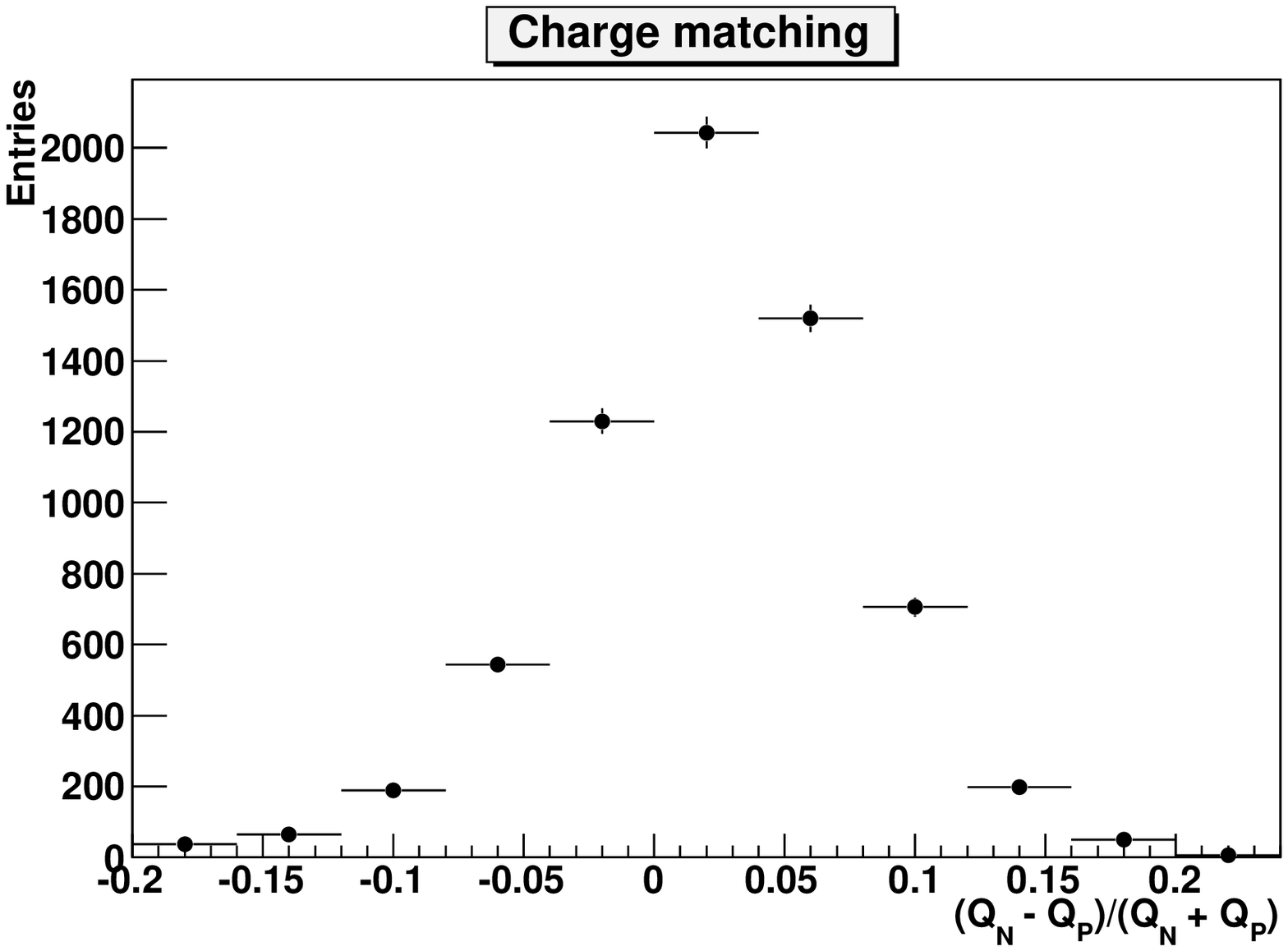} 
\includegraphics[width=3.9cm,height=6cm]{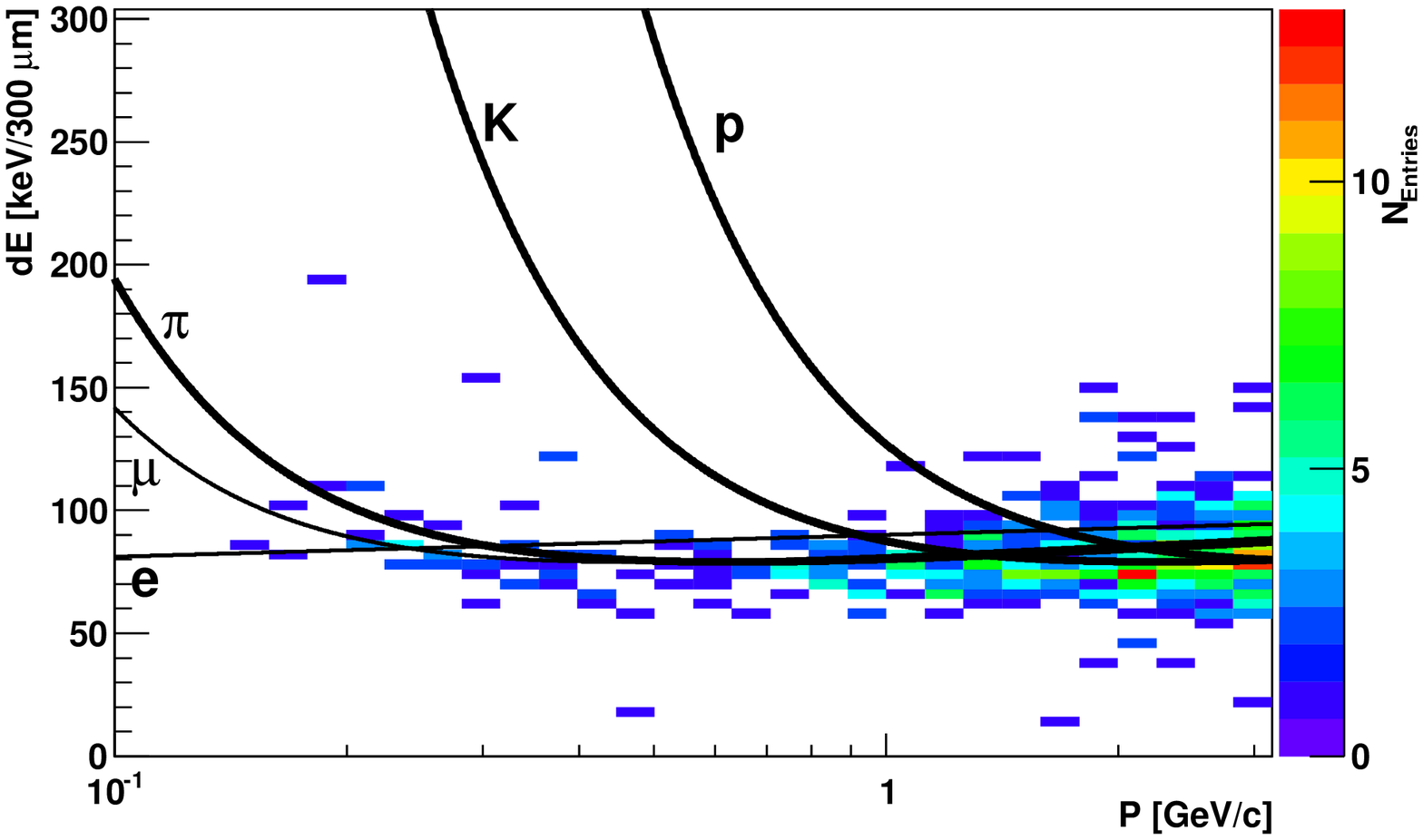} 
\includegraphics[width=3.9cm,height=6cm]{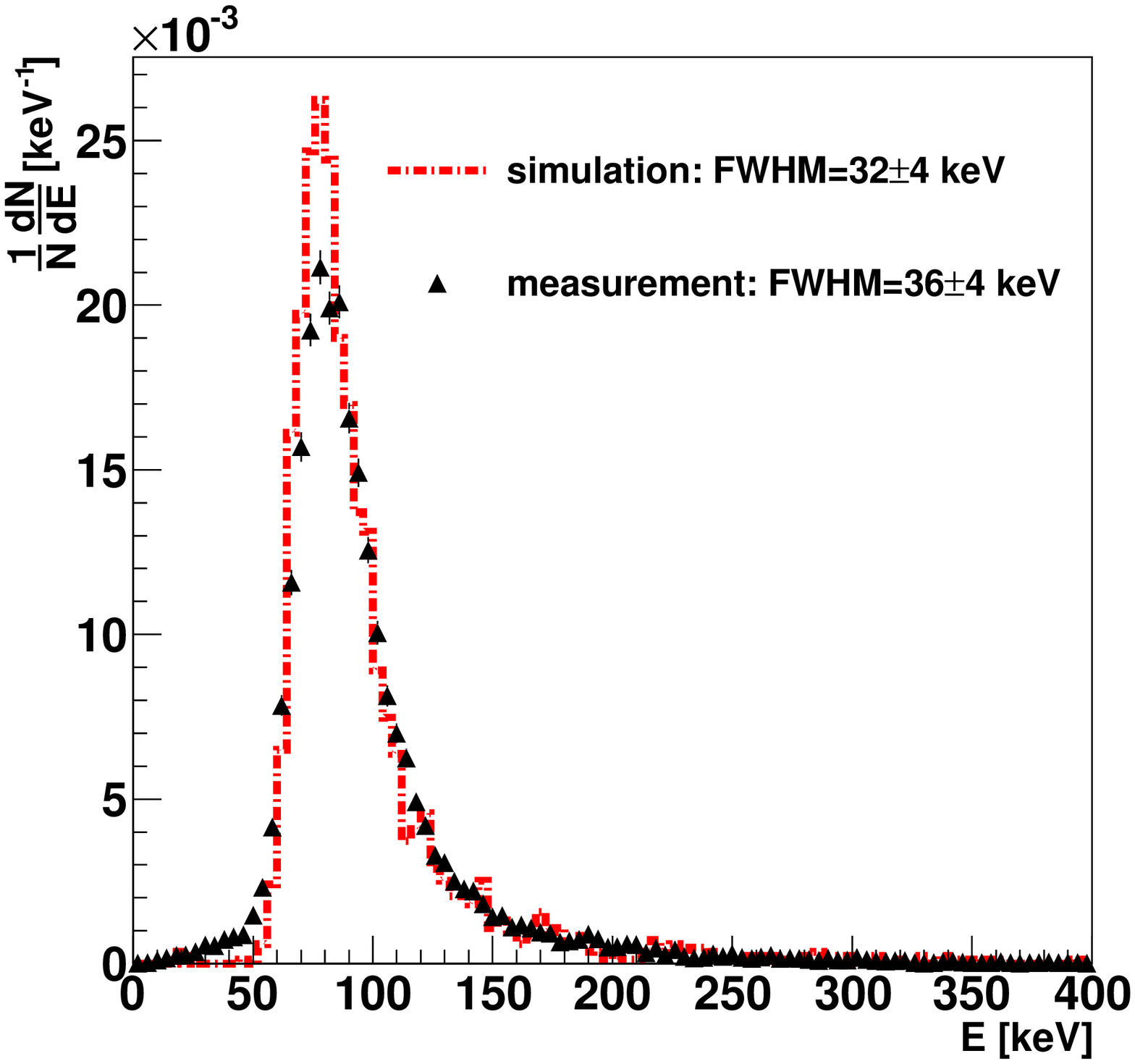} 
\caption{The distribution of the normalized difference in P- and N- charge 
(left). The dE/dx distributions as a function of the particle's momenta for 
cosmic data (middle). The charge distribution for both simulations and cosmic 
data (right).} 
\label{fig:calibration} 
\end{figure}

For the charge matching, we relied on runs without the presence of the magnetic 
field. Only modules with large number of accumulated cluster statistics were 
considered. Corrections were applied to get the best calibration results at 
the module level. The resulting normalized difference in P- and N-charge, as 
illustrated in fig. \ref{fig:calibration}-left, has a FWHM of $11 \%$ 
\cite{Ref:MarekNote}. Detailed studies were performed to check the calibration 
constants at the chip level though the accumulated statistics was not 
sufficient to extract a definite conclusion. Preliminary results show that the 
calibration at the chip level improves the stability of the P- and N-side gains 
by $10 \%$ \cite{Ref:MarekNote}.

For the absolute calibration, the analysis of cosmic data obtained with the 
magnetic field was used. Due to the fact that the TPC calibration was not 
optimized and resulted in a poor ITS-TPC track matching, we mainly relied 
on the stand-alone ITS tracking \cite{Ref:AlicePPR}. Corrections were applied 
to take into account the inclination of the track, thus addressing the issue 
of the different track lengths when a particle crosses a module.

Figure \ref{fig:calibration}-middle shows the energy loss measured by the two 
SSD layers as a function of the particles' momenta for the cosmic data sample 
analyzed. The curves are drawn based on \cite{Ref:SiEnergyLoss}. Figure 
\ref{fig:calibration}-right gives the comparison of the charge distributions 
for both simulated and real data. The FWHM obtained for the two distributions 
differ by $12.5 \%$.

\section{Summary}

In this paper we presented the main results from the commissioning of the SSD 
of ALICE using cosmic data. The SSD participated with great efficiency 
in the different ALICE commissioning periods started in December 2007. The 
successful cosmics data taking in summer 2008, allowed us to perform the first 
part of the detector's alignment and calibration. The data that are going to 
be collected in the upcoming period will be used for the further refinement of 
these two activities. In conclusion, the SSD performance results are very 
close to the designed specifications and is ready for the first pp and Pb-Pb 
LHC collisions.

\end{document}